\documentclass[useAMS,usenatbib]{mn2e}
\usepackage{times}
\usepackage{graphicx}
\usepackage{subfigure}
\usepackage{psfrag}
\usepackage{amsmath}
\usepackage{amssymb}
\usepackage{color}
\usepackage{mathrsfs}

\def\reff@jnl#1{{\rm#1\/}}
\def\aj{\reff@jnl{AJ}}                 
\def\araa{\reff@jnl{ARA\&A}}           
\def\apj{\reff@jnl{ApJ}}               
\def\apjl{\reff@jnl{ApJ}}              
\def\apjs{\reff@jnl{ApJS}}             
\def\ao{\reff@jnl{Appl.Optics}}        
\def\apss{\reff@jnl{Ap\&SS}}           
\def\aap{\reff@jnl{A\&A}}              
\def\aapr{\reff@jnl{A\&A~Rev.}}        
\def\aaps{\reff@jnl{A\&AS}}            
\def\azh{\reff@jnl{AZh}}               
\def\baas{\reff@jnl{BAAS}}             
\def\jcap{\reff@jnl{JCAP}}             
\def\jrasc{\reff@jnl{JRASC}}           
\def\memras{\reff@jnl{MmRAS}}          
\def\mnras{\reff@jnl{MNRAS}}           
\def\pra{\reff@jnl{Phys.Rev.A}}        
\def\prb{\reff@jnl{Phys.Rev.B}}        
\def\prc{\reff@jnl{Phys.Rev.C}}        
\def\prd{\reff@jnl{Phys.Rev.D}}        
\def\prl{\reff@jnl{Phys.Rev.Lett}}     
\def\pasp{\reff@jnl{PASP}}             
\def\pasj{\reff@jnl{PASJ}}             
\def\qjras{\reff@jnl{QJRAS}}           
\def\skytel{\reff@jnl{S\&T}}           
\def\solphys{\reff@jnl{Solar~Phys.}}   
\def\sovast{\reff@jnl{Soviet~Ast.}}    
\def\ssr{\reff@jnl{Space~Sci.Rev.}}    
\def\zap{\reff@jnl{ZAp}}               
\def\nat{\reff@jnl{Nature}}            

\title[Mutual consistency test of SNe surveys]
{Testing the mutual consistency of different supernovae surveys}
\author[N.V.~Karpenka, F.~Feroz and M.P.~Hobson]  
{N.V.~Karpenka$ ^{1} $\thanks{E-mail: nkarp@fysik.su.se}, F.~Feroz$ ^2 $ 
and M.P.~Hobson$ ^2 $\\ 
$^1$The Oskar Klein Centre for Cosmoparticle Physics, Department of Physics, Stockholm University, AlbaNova, SE-106 91 Stockholm, Sweden\\
$^2$Astrophysics Group, Cavendish Laboratory, J.J.~Thomson Avenue, Cambridge CB3 0HE, UK\\}

\date{Accepted ---. Received ---; in
  original form \today}

\pagerange{\pageref{firstpage}---\pageref{lastpage}}
\pubyear{2014}

\begin{document}
\label{firstpage}
\maketitle
\begin{abstract}
It is now common practice to constrain cosmological parameters using
supernovae (SNe) catalogues constructed from several different
surveys. Before performing such a joint analysis, however, one
should check that parameter constraints derived from the individual
SNe surveys that make up the catalogue are mutually consistent. We
describe a statistically-robust mutual consistency test, which we
calibrate using simulations, and apply it to each pairwise combination
of the surveys making up, respectively, the UNION2 catalogue and the
very recent JLA compilation by Betoule et al. We find no
inconsistencies in the latter case, but conclusive evidence for
inconsistency between some survey pairs in the UNION2 catalogue.
\end{abstract}


\section{Introduction}\label{sec:intro}

Cosmology using type Ia supernovae (SNIa) has become a very successful
and important field over the last 15 years, with SNIa playing a
central role in the discovery of the accelerated expansion of the
universe \citep{1998AJ....116.1009R, 1999ApJ...517..565P}. This
accelerated expansion is broadly interpreted as an important piece of
evidence for the existence of an exotic dark energy component. The
basic approach assumes that SNIa constitute a set of `standardizable'
candles, since by applying small corrections to their absolute
magnitudes, derived by fitting multi-wavelengths observations of their
light-curves, one can reduce the scatter magnitude considerably, to
around $\pm 0.15$~mag in the $B$-band (\citealt{
  1993ApJ...413L.105P,1996AJ....112.2391H}).  In essence, SNIa with
broader light-curves and slower decline rates are intrinsically
brighter than those with narrower light-curves and faster decline
rates, similarly, bluer SNIa are intrinsically brighter than their
redder counterparts \citep{1998A&A...331..815T}. Then, by measuring
the distance moduli and redshifts of a set of SNIa, and comparing
these with the predicted distance modulus for each SNIa, one can place
constraints on the cosmological parameters.

From the very first such studies, SNIa from different telescopes have
been combined in a joint analysis in order to have SNe both at low and
high $z$ \citep{1998AJ....116.1009R, 1999ApJ...517..565P}.  During the last few years, large combined SNe compilations (see e.g. \citealt{2012ApJ...746...85S}) are becoming widely used in SNe cosmology. Although such
analyses result in tighter constraints on parameters, one needs to be
careful when using data from different sources, as the presence of
unaccounted systematics in any of these data-sets can lead to
misleading results. Therefore, it is extremely important to establish
whether different data-sets are consistent with each other before
performing a joint analysis using them.

A common method to check for consistency between different data-sets,
in general, is to perform a $\chi^2$ analysis and compare the best-fit
values $\chi^2_{\rm min}$ obtained by using (i) all the data-sets
together and (ii) each data-set independently. If the data-sets are
mutually consistent, then one would expect that in each case
$\chi^2_{\rm min} \sim N_{\rm dof} \pm \sqrt{2N_{\rm dof}}$, where
$N_{\rm dof}$ is the number of degrees of freedom in the fit.  It is,
however, impossible to perform such a test in the standard framework
of SNe analysis, since the intrinsic dispersion $\sigma_{\rm int}$,
which describes the global variation in the SNIa absolute magnitudes
that remains after correction for stretch and colour, is always
adjusted so that a good fit is obtained (i.e. $\chi^2_{\rm min}
\approx N_{\rm dof}$).  Moreover, quite generally, tests based on
$\chi^2_{\rm min}$ depend only on the best-fit values of parameters,
and are insensitive to the likelihood over the rest of the parameter
space.

One must therefore seek an alternative test for investigating
consistency between different SNIa data-sets.  Recently,
\citet{2013MNRAS.tmp..694A,2014MNRAS.439.1855H}
carried out a search for systematic contamination within the UNION2.1
SNe catalogue \citep{2012ApJ...746...85S} using a method based on the
Bayesian consistency test described in
\cite{2006PhRvD..73f7302M}. They concluded that UNION2.1 data do not
suffer from significant systematic effects. Crucially, however, in
this study some of the (non-cosmological) nuisance parameters were
fixed, thereby eliminating by design any potential inconsistency
arising from them.

The aim of this paper is to check for consistency between different
SNe data-sets by using a statistically robust method, again based on
the Bayesian test described in \cite{2006PhRvD..73f7302M}, but making
as few assumptions as possible on the fitted parameters. This paper is
organized as follows. We describe our analysis methodology including
the details of our consistency test in Section \ref{sec:method}. A
short description of simulations and data we have used in this work is
given in Section \ref{sec:data}. Our results are given in Section
\ref{sec:res}. Finally, our conclusions are presented in Section
\ref{sec:conc}.

\section{Analysis methodology}\label{sec:method}

SNIa survey data consist of the apparent magnitudes or fluxes observed
in different filters, depending on the particular survey, at a series
of epochs. Estimation of cosmological parameters from these flux
measurements is a two-step procedure. First, a SN light-curve fitting
algorithms is applied to these raw data in order to `standardise'
these observations by reducing the scatter in the Hubble
diagram. Second, cosmological parameters are estimated using the
output from the light-curve fitting method. In our analysis, this is
followed by the application of a consistency test to the constraints
derived both on the cosmological parameters and those associated with
the SNe population. 

\subsection{Light-curve fitting}
\label{sec:lightcurvefit}

Standard algorithms for light-curve fitting include the
Spectral Adaptive Light-curve Template method (SALT)
\citep{GuyAstier2005,AstierGuy2006}, SALT-II \citep{GuySullivan2010},
SiFTO \citep{ConleySullivan2008}, Multi Colour light-curve Shape
(MLCS) method \citep{JhaRiess2007}, Gaussian Process Data Regression
\citep{2013ApJ...766...84K}, and Hierarchical Models
\citep{2011ApJ...731..120M}. In this paper we employ the SALT-II SN
light-curve fitter, since it was used to obtain the data-sets
described in Section~\ref{sec:data:real}.

When fitting SN light-curves with SALT-II, the outputs are the
best-fit values $\hat{m}_{B}^\ast$ of the apparent rest frame $B$-band
magnitudes of the SNe at maximum luminosity, the light-curve shape
parameter $\hat{x}_1$ related to the width (stretch) of the fitted
light-curve, the colour $\hat{c}$ in the $B$-band at maximum
luminosity, and the covariance matrix of the uncertainties in the
estimated light-curve parameters
\begin{equation}
\widehat{C} = 
\left(
\begin{array}{ccc}
\sigma^2_{m^\ast_B} & \sigma_{m^\ast_B,x_1} & \sigma_{m^\ast_B,c} \\[2mm]
\sigma_{m^\ast_B,x_1} & \sigma^2_{x_1} & \sigma_{x_1,c}\\[2mm]
\sigma_{m^\ast_B,c} & \sigma_{x_1,c} &  \sigma^2_{c} \\
\end{array}
\right).
\label{eq:covmat}
\end{equation}
Combining these quantities with the estimated redshift $\hat{z}$ of
the SN, our basic input data for each SN are thus
\begin{equation}
D_i \equiv \{\hat{z}_i,\hat{m}_{B,i}^\ast,\hat{x}_{1,i},\hat{c}_i\}
\label{eqn:inputdata}
\end{equation}
for ($i=1,\ldots,N_{\rm SN}$). The vector of values
$(\hat{m}_{B,i}^\ast, \hat{x}_{1,i}, \hat{c}_i)$ for each SN is then
assumed to be distributed as a multivariate Gaussian about their
respective true values, with covariance matrix $\widehat{C}_i$.

\subsection{Parameter estimation}

The output from the light-curve fitting constitutes the data used to
estimate parameters, both cosmological and those associated with the
SNe population. The usual way in which this analysis is performed by
the SNe community is to use the standard $\chi^2$-method, see for example
\cite{AstierGuy2006,Kessler2009Firstyear,GuySullivan2010}.  Recently, other inference methods have been proposed, such as
a likelihood approach by \citet{2012A&A...541A.110L} and the Bayesian
Hierarchical Method (BHM; \citealt{2011MNRAS.418.2308M,
  2012arXiv1207.3705M}), both of which are better motivated
statistically than the standard $\chi^2$ analysis. Nonetheless, in
this paper we employ the $\chi^2$ method, since it remains the
approach most commonly used by the SNe community.

In this approach, the $\chi^2$ misfit function is defined to be
\begin{equation} 
\label{eq:chisq}
\chi^2(\mathscr{C},\alpha,\beta, M_0,\sigma_\text{int}) = \sum_{i=1}^N 
\frac{[\mu_i^\text{obs}(\alpha,\beta,M_0)-\mu(\hat{z}_i,\mathscr{C})]^2}{\sigma_i^2(\alpha,\beta,\sigma_\text{int})},
\end{equation}
where, for clarity, we have made explicit the functional dependencies
of the various terms on (only) the parameters to be fitted. In this
expression, the `observed' distance modulus for the $i$th SN is
\begin{equation}
\mu_i^{\rm obs} = \hat{m}_{B,i}^\ast - M_0 + \alpha \hat{x}_{1,i} - \beta\hat{c}_i,
\label{eq:muobsdef}
\end{equation}
where $M_0$ is the (unknown) $B$-band absolute magnitude of the SNe,
and $\alpha$, $\beta$ are (unknown) nuisance parameters controlling
the stretch and colour corrections; all three parameters are assumed
to be global, i.e. the same for all SNIa. The predicted distance modulus
is given by
\begin{equation}
\mu(z,\mathscr{C}) =
5\log_{10}\left[\frac{D_{L}(z,\mathscr{C})}{\text{Mpc}}\right]+25,
\label{pred-dist-mod}
\end{equation}
where $D_{L}(z,\mathscr{C})$ is the luminosity distance to an object
of redshift $z$ in a universe with cosmological parameters
$\mathscr{C}= \{\Omega_{m,0}, \Omega_{\Lambda,0}, H_0, w\}$. The total
dispersion $\sigma_{i}^2$ is the sum of several errors added in
quadrature:
\begin{equation}
\sigma_{i}^2 = (\sigma_{\mu,i}^z)^2 + \sigma_\text{int}^2 
+ \sigma_{\text{fit},i}^2(\alpha,\beta).
\label{eq:sigmadef}
\end{equation}
The three components are: (i) the error $\sigma_{\mu,i}^z$ in the
distance modulus that is induced by the error $\sigma_{z,i}$ in the redshift
measurement, which may be due to uncertainties in the peculiar velocity of the
host galaxy or SNIa and in the spectroscopic measurements of either the host galaxy of the SNIa itself; (ii) the intrinsic
dispersion $\sigma_\text{int}$, which describes the global variation
in the SNIa absolute magnitudes that remain after correction for
stretch and colour, which quantifies all of the unknown intrinsic
dispersion errors; and (iii) the fitting error, which is given by
\begin{equation}
\sigma_{\text{fit},i}^2(\alpha,\beta) = \bpsi^{\rm t} \widehat{C}_i  \bpsi,
\label{eq:sigmafitdef}
\end{equation}
where the transposed vector $\bpsi^{\rm t} = \left(1, \alpha, -\beta
\right)$ and $\widehat{C}_i$ is the covariance matrix given in
(\ref{eq:covmat}).

Typically, the chi-squared function (\ref{eq:chisq}) is minimized
simultaneously with respect to the cosmological parameters
$\mathscr{C}$ and the global SNIa nuisance parameters $M_0$, $\alpha$
and $\beta$, with $\sigma_\text{int}$ fixed to some initial estimate.
There are, however, a few differences in the way in which this
minimisation is performed, such as which search algorithm is used
(MCMC techniques or grid searches) and the treatment of $M_0$ (which
is degenerate with $H_0$), in particular whether this parameter is
marginalised over analytically or numerically. Once the minimum value
of chi-squared has been obtained, the value of $\sigma_\text{int}$ is
updated by adjusting it to obtain $\chi^2_{\rm min}/N_{\rm dof} \sim
1$; the entire process may be iterated, if necessary, until
$\sigma_\text{int}$ converges.

In this paper, we accommodate the exact degeneracy between $M_0$ and
$H_0$ by fixing $H_0=73$~km~s$^{-1}$~Mpc$^{-1}$, which is consistent
with independent constraints from other probes \cite{2013arXiv1303.5062P}, and
allow $M_0$ to vary. For simplicity, we also assume that dark energy
is in the form of a cosmological constant $(w=-1)$ and that the
universe is spatially-flat
$(\Omega_{\Lambda,0}+\Omega_{\text{m},0}=1)$; thus we vary only the
parameters
$\{\Omega_{\text{m},0},M_0,\alpha,\beta,\sigma_\text{int}\}$.  In
keeping with standard practice, we define the `likelihood' to be
simply
\begin{equation}
\mathcal{L}(\Omega_{\text{m},0}, M_0, \alpha,
\beta,\sigma_\text{int}) = \exp
     [-\tfrac{1}{2}\chi^2(\Omega_{\text{m},0}, M_0, \alpha, \beta,
       \sigma_\text{int})]
\label{eq:like}
\end{equation}
which is clearly not properly normalized.  More importantly, this
`likelihood' is {\em not} (proportional to) the probability
$\Pr(\bmath{D}|\Omega_{\text{m},0}, M_0, \alpha,
\beta,\sigma_\text{int})$. Statistically well-motivated expressions
for the latter may be found using the likelihood or BHM approaches
mentioned above, but we will not pursue such methods here.  

Adopting the initial value $\sigma_\text{int}=0.1$, we
use the nested sampling algorithm {\sc MultiNest} \citep{2008MNRAS.384..449F, 2009MNRAS.398.1601F, 2013arXiv1306.2144F} to
sample from the (unnormalised) `posterior'
$\mathcal{P}(\mathbf{\Theta},\sigma_\text{int}) =
\mathcal{L}(\mathbf{\Theta},\sigma_\text{int})\pi(\mathbf{\Theta})$
over the parameter space
$\mathbf{\Theta}=\{\Omega_{\text{m},0},M_0,\alpha,\beta\}$, assuming
the separable, uniform priors $\pi(\mathbf{\Theta})$ listed in
Table~\ref{tab:priors}. Following the standard approach, the value of
$\sigma_\text{int}$ is estimated by adjusting it to obtain
$\chi^2_{\rm min}/N_{\rm dof} \sim 1$, where $\chi^2_{\rm min}$ is the
value at the maximum $\widehat{\mathbf{\Theta}}$ of the `posterior'
(and also of the `likelihood', since the assumed priors are
uniform). The sampling over the parameter space $\mathbf{\Theta}$ is
then repeated. We find that one or two such iteration on $\sigma_\text{int}$
is sufficient to achieve convergence.
\begin{table}
\begin{center}
\begin{tabular}{lccr}
\hline
Parameter & Symbol & Prior & \hspace*{-0.3cm}Simulated value \\
\hline
Matter density & $\Omega_{\text{m},0}$ & ${\cal U}(0,1)$ & 0.3\\
Absolute magnitude & $M_0$ & ${\cal U}(-20.3,-18.3)$ & $-19.3$\\
Stretch multiplier & $\alpha$ & ${\cal U}(0,1)$ & 0.12 \\
Colour multiplier & $\beta$ & ${\cal U}(1,5)$& 2.6 \\
\hline
\end{tabular}
\caption{Priors assumed on the parameters, where $U(a,b)$ indicates
  the uniform distribution in the range $[a,b]$. The final column
  gives the value assumed in generating the simulated data.\label{tab:priors}}
\end{center}
\end{table}

We note that, when jointly analysing a full catalogue of SNIa data,
$\sigma_\text{int}$ is the same for all SNe, regardless of whether
they belong to the same survey or not. Apart from modelling the
intrinsic dispersion in distance moduli of SNe, however,
$\sigma_\text{int}$ can also account for systematics in different
surveys. Therefore different surveys are allowed to have different
values of $\sigma_\text{int}$ when analysing them separately.

\subsection{Test for mutual consistency between data-sets}

The standard approach to analysing multiple data-sets jointly is
simply to assume that they are mutually consistent. We represent this
(null) hypothesis by $H_{0}$. This is the approach adopted in the vast
majority of analyses in SNe cosmology. It may be the case, however,
that the data-sets are inconsistent with one another, resulting in
each one favouring a different region of the model parameter space. We
represent this (alternative) hypothesis by $H_{1}$. In this case, a
joint analysis would lead to completely misleading results (see, for
example, Appendix A in \citealt{2008JHEP...10..064F} for a
demonstration).

In order to determine which one of these hypotheses is favoured by the
data, one can perform Bayesian model selection between $H_{0}$ and
$H_{1}$. Assuming that hypothesis $H_0$ and $H_1$ are equally likely
{\em a priori}, this can be achieved by calculating the ratio
\begin{equation}
\mathcal{R} = \frac{\Pr(\mathbf{D}|H_{0})}{\Pr(\mathbf{D}|H_{1})}
  = \frac{\Pr(\mathbf{D}|H_{0})}{\prod_{i}\Pr(D_{\rm i}|H_{1})}.
\label{eq:R-test}
\end{equation}
where the probabilities $\Pr(\mathbf{D}|H)$, called Bayesian evidences
or marginal likelihoods, are defined as
\begin{equation}
\Pr(\mathbf{D}|H) = \int{\Pr(\mathbf{D}|\mathbf{\Theta}, H)\Pr(\mathbf{\Theta}|H)}\mathrm{d}^n\mathbf{\Theta}.
\label{eq:evid}
\end{equation} 
Here $n$ is the dimensionality of the parameter space,
$\Pr(\mathbf{D}|\mathbf{\Theta}, H)$ is the likelihood and
$\Pr(\mathbf{\Theta}|H)$ is the prior. The numerator of
(\ref{eq:R-test}) represents the standard joint analysis of all the
data-sets $\mathbf{D} = \{D_1, D_2, \cdots, D_n\}$, whereas the
denominator in the final expression represents the case in which each
data-set is analysed separately It is worth noting, however, that the
second equality in (\ref{eq:R-test}) is valid only when one allows for
potential inconsistencies in the preferred values of the {\em full}
set of model parameters $\mathbf{\Theta}$.

In adapting this Bayesian test to apply it to the standard $\chi^2$
analysis of SNIa data, $\Pr(\mathbf{D}|\mathbf{\Theta}, H)$ is replaced by the `likelihood'
$\mathcal{L}(\mathbf{\Theta},\sigma_\text{int})$ given in
(\ref{eq:like}) and $\Pr(\mathbf{\Theta}|H)$ by the prior
$\pi(\mathbf{\Theta})$ listed in Table~\ref{tab:priors}. In so doing,
however, the terms in (\ref{eq:R-test}) can not be interpreted directly
as probabilities, and thus the value of the $\mathcal{R}$ cannot be
compared with the normal Jeffrey's scale for model selection based on
the ratio of evidences. Nonetheless, $\mathcal{R}$
values are still expected to be higher for consistent data-sets and lower for
inconsistent ones \citep{2011MNRAS.415..143M}, so $\mathcal{R}$ may be
used as a test statistic that is `calibrated' with simulations to
perform a standard one-sided frequentist hypothesis test. 

Thus, we construct the distribution of $\mathcal{R}$ under the null
hypothesis $H_0$ by evaluating it for a large set of simulations in
which the individual surveys making up a catalogue are mutually
consistent. The $\mathcal{R}$ value obtained by analysing the real
data can then be used to calculate the $p$-value as follows:
\begin{equation}
 p = \frac{N(\mathcal{R}_{\rm s} < \mathcal{R}_{\rm r})}{N_{\rm tot}},
\label{eq:p-value}
\end{equation}
where $\mathcal{R}_{\rm s}$ and $\mathcal{R}_{\rm r}$ are the
$\mathcal{R}$ values obtained by analysing simulated and real
data-sets respectively, $N(\mathcal{R}_{\rm s} < \mathcal{R}_{\rm r})$
is the number of simulations with $\mathcal{R}$ values less than that
obtained by analysing the real data and $N_{\rm tot}$ is the total
number of simulations. We use the scale given in
Table~\ref{tab:thresholds} to draw inferences from these $p$-values
about the mutual consistency of the data-sets
\begin{table}
\begin{center}
\begin{tabular}{cc}
\hline
$p$-value & Conclusion \\
\hline
$< 0.01$ & very strong presumption against $H_0$  \\
$ 0.01 - 0.05$ & strong presumption against $H_0$   \\
$ 0.05 - 0.10$ & low presumption against $H_0$  \\
$> 0.10$ &  no presumption against $H_0$ \\[1mm]
\hline
\end{tabular}
\caption{Thresholds for the $p$-value.}
\label{tab:thresholds}
\end{center}
\end{table}
%

\section{Data-sets}\label{sec:data}
\subsection{Real data}\label{sec:data:real}

In this paper we consider SNe from the UNION2 catalogue
\citep{2010ApJ...716..712A} and the `joint light-curve analysis' (JLA)
compilation given in \cite{2014arXiv1401.4064B}.  These data-sets
consist of SNe observed by different telescopes. Since our aim in this
paper is to check for consistency between different SNe
surveys/data-sets, we divide these SNe according to the telescope with
which they have been observed. In particular, we divide SNe into the
following subsets: ESSENCE \citep{2007ApJ...666..674M,
  2007ApJ...666..694W, 2005AJ....130.2453K}, Hubble Space Telescope
(HST; \citealt{1998ApJ...493L..53G, 2003ApJ...598..102K,
  2004ApJ...607..665R, 2007ApJ...659...98R}), Sloan Digital Sky Survey
(SDSS; \citealt{2008AJ....135..338F, K09, 2008AJ....136.2306H,2014arXiv1401.4064B}),
Supernova Legacy Survey (SNLS; \citealt{2006A&A...447...31A,
  2010A&A...523A...7G} , S11), CfA (\citealt{2009ApJ...700..331H}) and
a compilation of nearby SNIa measurements
(\citealt{1996AJ....112.2408H, 1999AJ....117..707R,
  2010AJ....139..519C, 2006AJ....131..527J}). Even though the SNe in
the nearby compilation have not all been observed from the same
telescope, we treat them as if they were. This is normal practice in
SN analyses, and is justified since such SNe are usually observed with
similar precision and sets of filters. The resulting number of SNe in
the UNION2 and JLA compilations coming from each of the subsets is
given in Table~\ref{tab:numberSNe}. These corresponds to all the SNe
in the JLA compilation, but does exclude some SNe from the UNION2
catalogue, which were observed by telescopes contributing just a
handful of SNe in each case.
\begin{table}
\begin{center}
\begin{tabular}{lrr}
\hline
Data-set & UNION2 & JLA \\
\hline
\phantom{(}ESSENCE & 74 &  ---\phantom{)} \\
\phantom{(}HST & 42  & 9\phantom{)}\\
\phantom{(}SDSS & 131  & 374\phantom{)}\\
\phantom{(}SNLS & 72 & 239\phantom{)}\\
\phantom{(}CfA & 100& ---\phantom{)} \\
\phantom{(}Low-$z$ & ---  & 118\phantom{)} \\
$\left(\mbox{Total}\right.$ & 419 & $\left. 740\right)$ \\[0.3mm]
\hline
\end{tabular}
\caption{Number of SNe in the UNION2 and JLA compilations coming from
  each survey. Note that the CfA survey in the UNION2 compilation and
  the low-$z$ survey in the JLA compilation have 51 SNe in common.}
\label{tab:numberSNe}
\end{center}
\end{table}

We use the publically-available fitted values of $\hat{m}_{B}^\ast$,
$\hat{x}_1$, $\hat{c}$ and $\widehat{C}$ from SALT-II light-curve fits
and the estimated redshifts $\hat{z}$ listed in the UNION2 and JLA
catalogues. For the JLA catalogue, however, we note that the quoted
covariance matrix $\widehat{C}$, in particular the element
$\sigma^2_{m_B^\ast}$, also contains the three variances listed on the
right-hand side of (\ref{eq:sigmadef}). To treat the UNION2 and JLA
catalogues in a consistent manner, we therefore subtracted these
contributions from $\sigma^2_{m_B^\ast}$ before analysing the JLA
catalogue.  We do not apply any additional cuts to the ones that have
already been applied to these catalogues. We refer readers to papers
describing these catalogues \cite{2010ApJ...716..712A,2014arXiv1401.4064B} for details on how the SNe have
been corrected for galactic extinction, peculiar velocities (at low
z), Malmquist and other selection biases.

\subsection{SNANA simulations}\label{sec:data:sim}

In order to calibrate our test statistic $\mathcal{R}$ and calculate the
resultant $p$-values, as described in Sec.~\ref{sec:method}, we
simulate SNe using the publicly available SNANA package\footnote{The SNANA
  package is publically-available and may be obtained from
  \texttt{http://sdssdp62.fnal.gov/sdsssn/SNANA-PUBLIC/}}
\citep{Kessler2009SNANA}. SNANA has a two-step simulation
process. First, data were simulated using appropriate light-curve
simulation templates to match the real data-sets closely.
The second stage is the light-curve fitting process in which the
photometric data are fitted to SALT II templates to give estimates for
$D_i \equiv
\{\hat{z}_i,\hat{m}_{B,i}^\ast,\hat{x}_{1,i},\hat{c}_i\}$. At this
light-curve fitting stage, basic cuts are made to discard SNIa with a
low signal-to-noise ratio and/or too few observed epochs in sufficient
bands. After the light-curve fitting stage we make a redshift
dependent magnitude correction for the Malmquist bias; the correction
is taken from a spline interpolation of Table 4 in
\cite{PerrettBalam2010}. All selection cuts and Malmquist corrections
are made prior to the cosmology inference step.

\begin{figure}
\centering
\includegraphics[width=\linewidth]{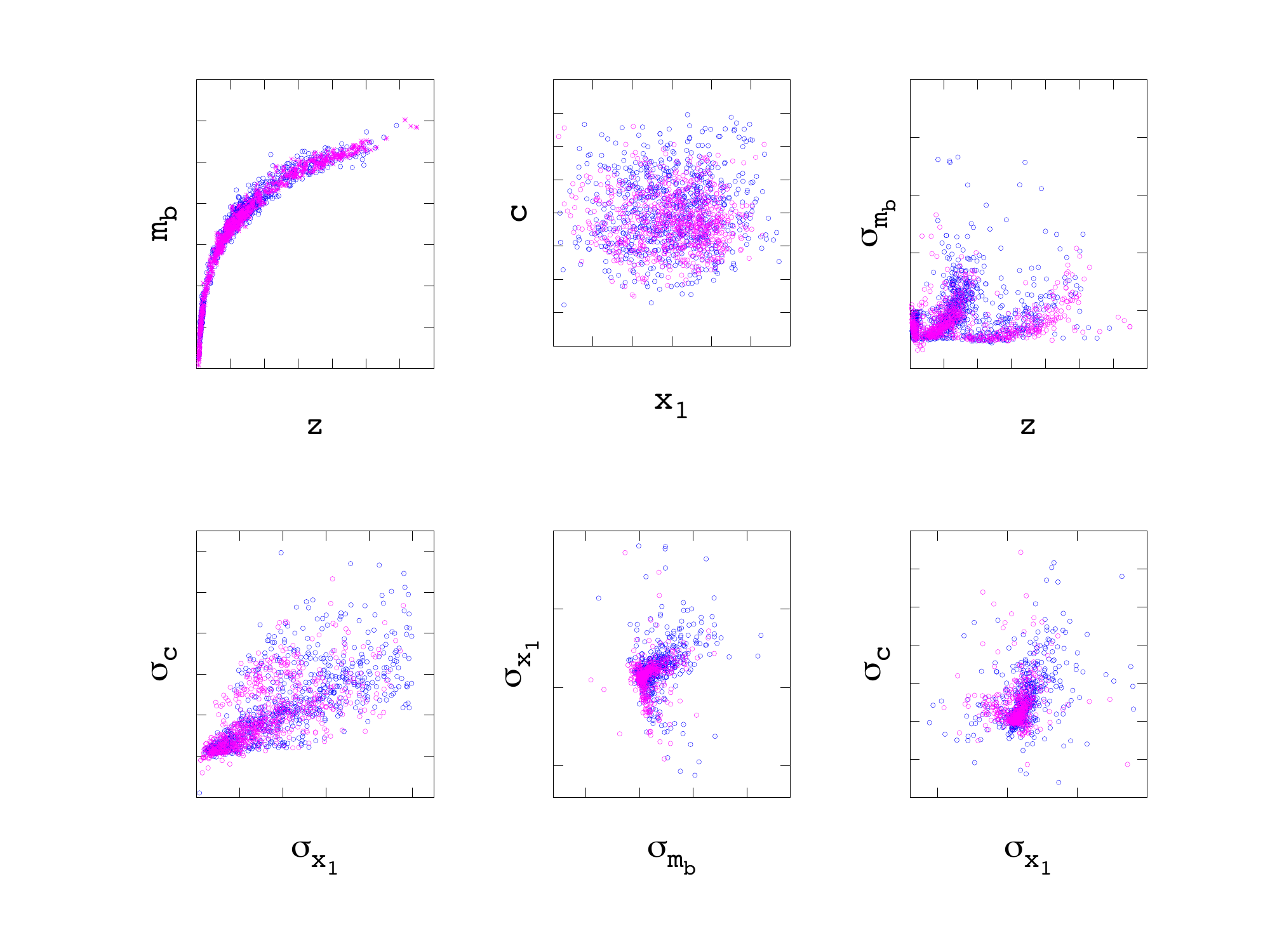} 
\caption{A simulated SNANA data-set (blue points) and real data from the JLA 
compilation (pink points).}
\label{fig:simVSdata}
\end{figure}

Our null hypothesis $H_0$ is that different data-sets are consistent
with each other. Therefore, we simulate all data-sets with exactly the
same parameter values, which are listed in Table \ref{tab:priors}, and
an intrinsic dispersion $\sigma_\text{int}= 0.106$. For the UNION2 and
JLA compilations, respectively, 100 data-sets were simulated for each
telescope/survey, with the numbers of SNe matching those listed in
Table~\ref{tab:numberSNe}. Fig. \ref{fig:simVSdata} shows a
compilation of one set of simulations, chosen at random, for each
telescope in the JLA compilation, overplotted on the real JLA data.  A
similar level of agreement between simulations and real data is
obtained for the UNION2 catalogue.

%
\section{Results}\label{sec:res}

\subsection{Constraints on $\Omega_{\rm m,0}$ and $M_0$}

\begin{figure}
\centering
\includegraphics[width=0.95\linewidth]{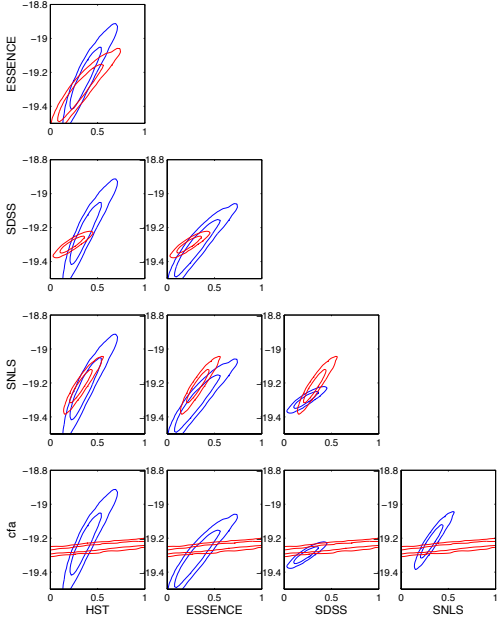} 
\caption{Two-dimensional marginalised constraints on the parameters
  $(\Omega_{\rm m,0},M_0)$ obtained from the individual constituent
  surveys contained in the UNION2 catalogue. The red (blue) contours
  denote the 68 and 95 per cent confidence regions for the survey in
  that row (column).\label{fig:U2cosmoconstraints}}
\end{figure}

\begin{figure}
\centering
\includegraphics[width=0.95\linewidth]{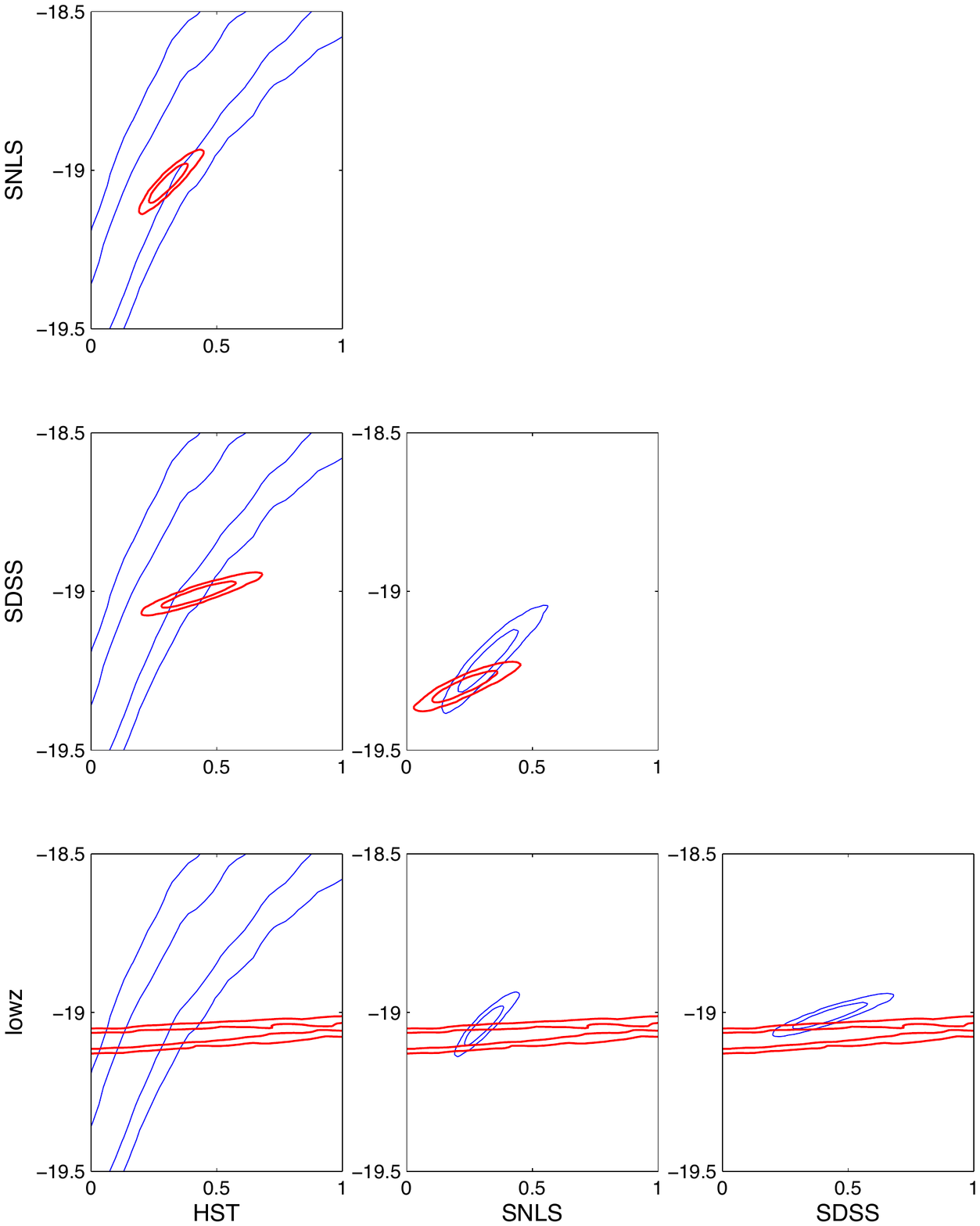} 
\caption{As in Fig.~\ref{fig:U2cosmoconstraints}, but for the JLA compilation.
\label{fig:JLAcosmoconstraints}}
\end{figure}

In analysing compilations of SNe data, it is usual simply to plot the
combined constraints on cosmological parameters obtained from a full
joint analysis. Here we begin by instead calculating the constraints
imposed by each constituent survey in the UNION2 and JLA compilations,
respectively. The resulting two-dimensional marginalised constraints
on the parameters $(\Omega_{\rm m,0},M_0)$ are shown in
Figs~\ref{fig:U2cosmoconstraints} and
\ref{fig:JLAcosmoconstraints}. We recall that $M_0$ is exactly
degenerate with $H_0$, and so we have fixed the latter to $H_0=73$ km
s$^{-1}$ Mpc$^{-1}$, in accordance with values obtained from
independent probes.
\begin{figure}
\centering
\includegraphics[width=0.95\linewidth]{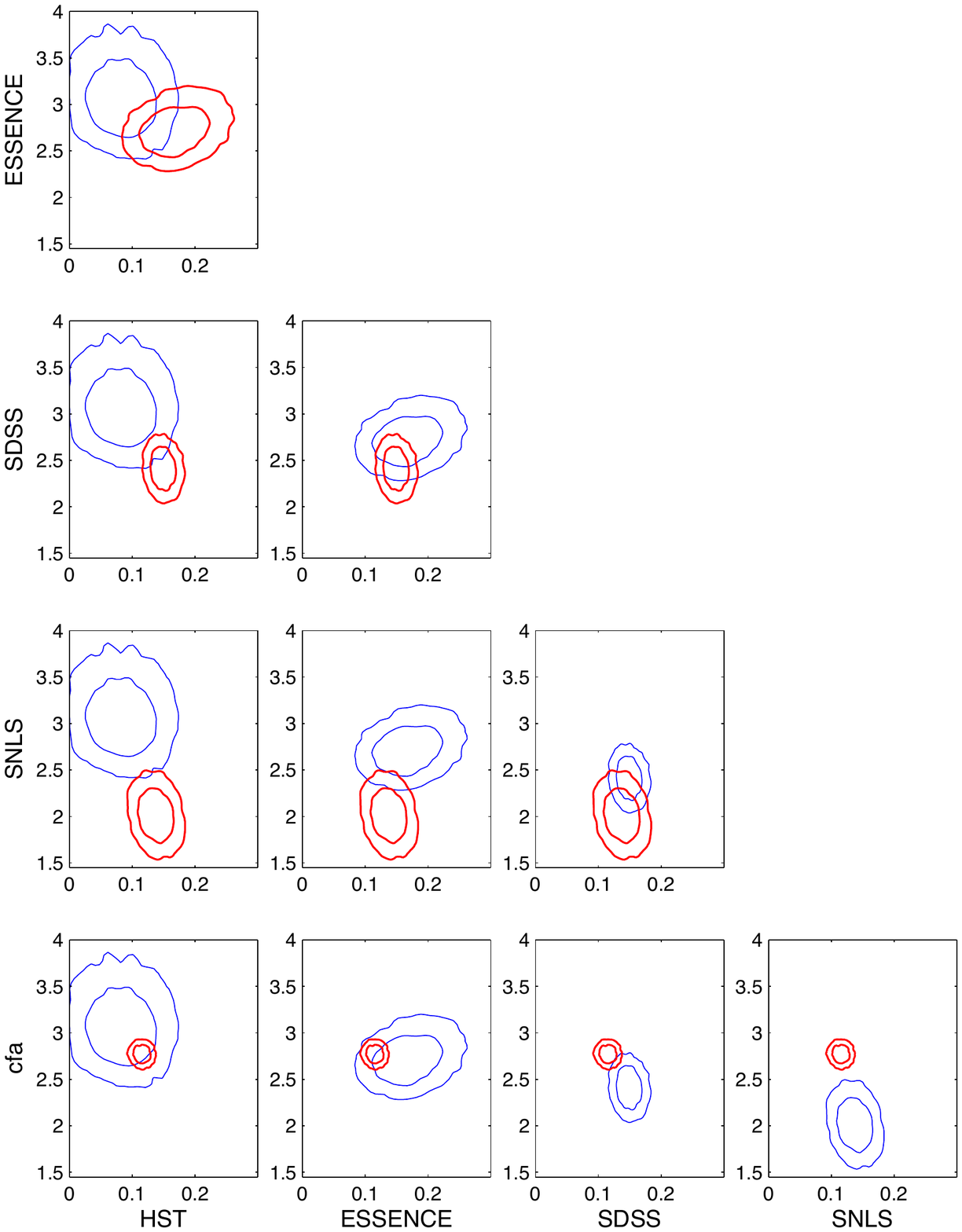} 
\caption{Two-dimensional marginalised constraints on the parameters
  $(\alpha,\beta)$ obtained from the individual constituent
  surveys contained in the UNION2 catalogue. The red (blue) contours
  denote the 68 and 95 per cent confidence regions for the survey in
  that row (column).\label{fig:U2abconstraints}}
\end{figure}
\begin{figure}
\centering
\includegraphics[width=0.95\linewidth]{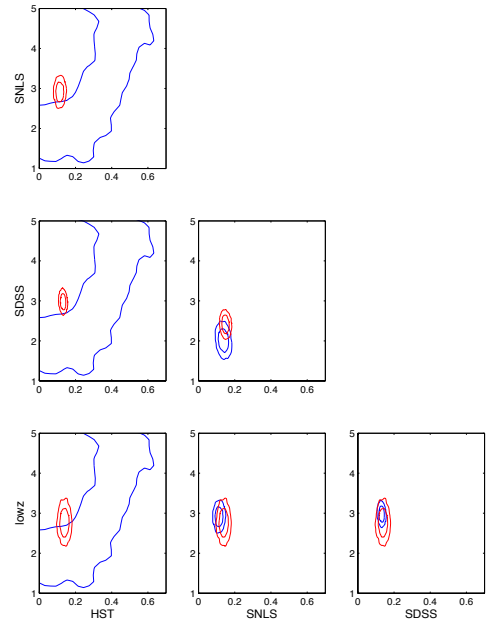} 
\caption{As in Fig.~\ref{fig:U2abconstraints}, but for the JLA compilation.
\label{fig:JLAabconstraints}}
\end{figure}

For the UNION2 compilation, the tightest constraints are
produced by the SDSS and SNLS surveys, with ESSENCE providing somewhat
weaker constraints. As might be expected, the CfA survey alone
constrains the parameters rather poorly, since it consists only of
nearby SNe. For the JLA compilation, the SDSS and
SNLS surveys again provide the tightest constraints, with the HST and
low-$z$ surveys constraining the parameters only very weakly, as one
would expect. For both compilations, however, the key observation for
our purposes is that the constraints from the constituent surveys are
in good agreement, as indicated by the large degree of overlap of the
confidence contours for each pairwise comparison.

\subsection{Constraints on $\alpha$ and $\beta$}

We now consider the constraints on the stretch and colour corrections
multipliers $\alpha$ and $\beta$, which are not often presented in the
analysis of SNe data. Once again we calculate the constraints
imposed by each constituent survey in the UNION2 and JLA compilations,
respectively. The resulting two-dimensional marginalised constraints
on the parameters $(\alpha,\beta)$ are shown in
Figs~\ref{fig:U2abconstraints} and \ref{fig:JLAabconstraints}.

For the UNION2 compilation, the tightest constraints are
produced by the CfA survey, with the remaining surveys yielding weaker
constraints all of a similar quality. It can be seen, however, that for
several pairings the confidence contours for the two surveys are
significantly displaced from one another, and do not overlap at all in
some cases.  This indicates mutual inconsistency between the survey
pairs and is most notable for CfA/SNLS, CfA/SDSS, SNLS/HST,
SNLS/ESSENCE and SDSS/HST. For the JLA compilation, the low-$z$, SDSS
and SNLS surveys all provide constraints of a similar quality, whereas
the HST constraints are very weak, as one might expect since this
corresponds to just 9 SNe. The key observation, however, is that for
each pairing the confidence contours for the two surveys overlap well,
indicating mutual consistency, in sharp contrast to the UNION2
results.

\begin{figure*}
\centering
\includegraphics[width=0.80\linewidth]{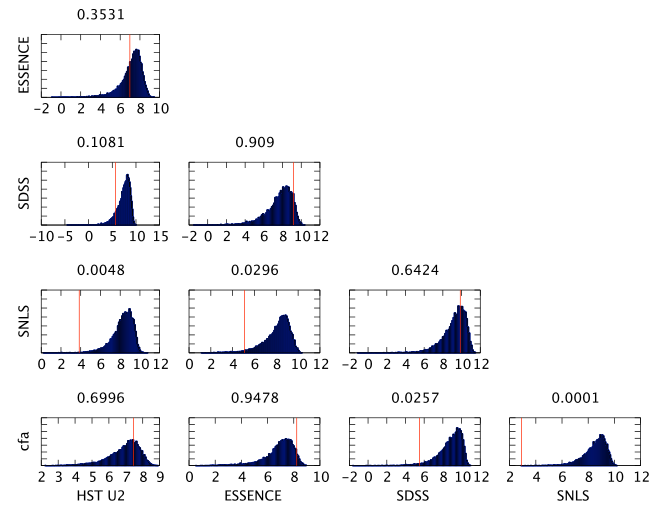} 
\caption{Results of the consistency test applied to surveys pairs in
  the UNION2 compilation. The blue histograms show the distribution of
  $R$ values obtained from $10^4$ consistent simulations of each pair,
  and the red vertical line indicates the $R$ value obtained from the
  real data. The corresponding one-sided $p$-value is given above each
  panel.
\label{fig:histU2}}
\end{figure*}
\begin{figure*}
\centering
\includegraphics[width=0.70\linewidth]{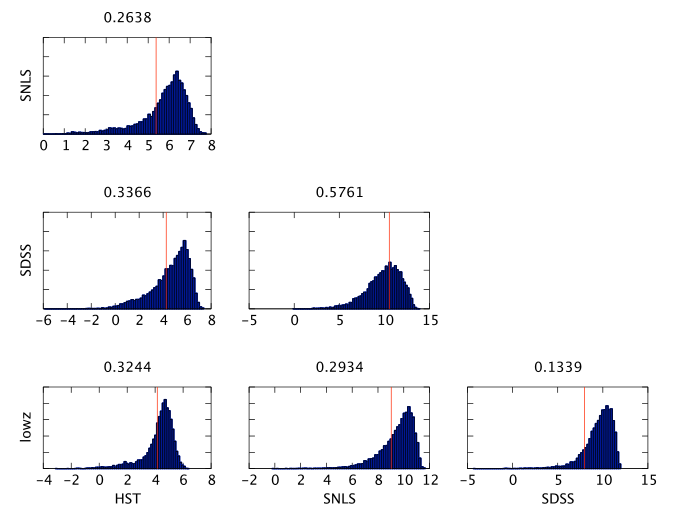} 
\caption{As in Fig.~\ref{fig:histU2}, but for the JLA compilation.
\label{fig:histNEW}}
\end{figure*}

\subsection{Mutual consistency test between survey pairs}
\label{sec:mutconpairs}

To quantify the level of mutual (in)consistency between survey pairs
in the UNION2 and JLA compilations, respectively, we calculate our test
statistic $\mathcal{R}$ for each combination, and compare it with the
distribution of $\mathcal{R}$ values for the corresponding survey pair
obtained from our consistent simulations. Since we generated 100
simulations for each survey, we have $10^4$ simulated pairs
in each case.

The results of the consistency test for the UNION2 compilation are
shown in Fig. \ref{fig:histU2}, together with the corresponding
$p$-values.
It is clear that the $p$-values obtained agree very closely with the
degree of overlap between the confidence contours plotted in
Fig.~\ref{fig:U2abconstraints}. In particular, we find very strong
evidence, at more than the 99 per cent significance level, for
inconsistency between the survey pairs CfA/SNLS and SNLS/HST. We also
find strong evidence, at more than 95 per cent significance, for
inconsistency in the pairs CfA/SDSS and SNLS/ESSENCE.

The results obtained for the JLA compilation are shown in
Fig.~\ref{fig:histNEW}
It can be seen that none of the survey pairs shows evidence for
inconsistency, even at the 90 per cent significance level, which is in
agreement with the good degree of overlap between the confidence
contours plotted in Fig.~\ref{fig:JLAabconstraints}. 

The consistency between the survey pairs in the JLA compilation is in
sharp contrast to inconsistencies present the UNION2 compilation. This
difference must result from the different SNe selection criteria and
calibration techniques used in the construction of the two
compilations.

\subsection{Effect of selection cuts on mutual consistency}

We now investigate the effect of selection cuts on two of the most
mutually inconsistent survey pairs within the UNION2 compilation,
namely CfA/SNLS and CfA/SDSS. To this end, we divide the 100 SNe in
the CfA data-set contained within UNION2 into two subsets of equal
size. The first subset (CfA-1) consists of 50 SNe chosen at random
from the 51 that are also present within the low-$z$ survey contained
in the JLA catalogue, and the second subset (CfA-2) consists of the
remaining 50 SNe in the CfA data-set. Clearly, all but one of the SNe in
CfA-2 are not present in the JLA catalogue. We then quantify the level
of mutual (in)consistency between each of these subsets and the SNLS
and SDSS surveys, respectively, in precisely the same manner as in
Section~\ref{sec:mutconpairs}.

\begin{figure}
\centering
\includegraphics[width=0.80\linewidth]{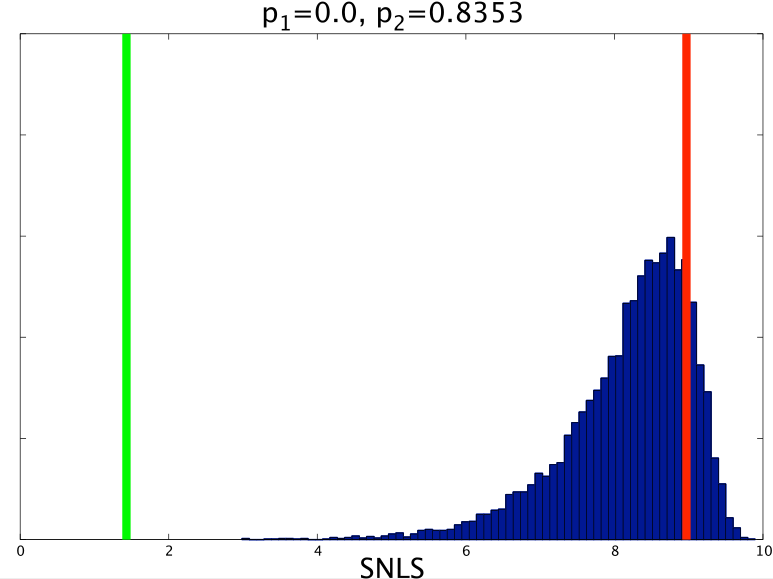} 
\caption{Results of the consistency test applied to the pairs
  CfA-1/SNLS (red vertical line) and CfA-2/SNLS (green vertical line)
  from the UNION2 compilation. The blue histogram shows the
  distribution of $R$ values obtained from consistent simulations of
  the (sub)surveys. The corresponding one-sided $p$-values
  are also given.\label{fig:snls-2sets}}
\end{figure}

\begin{figure}
\centering \includegraphics[width=0.80\linewidth]{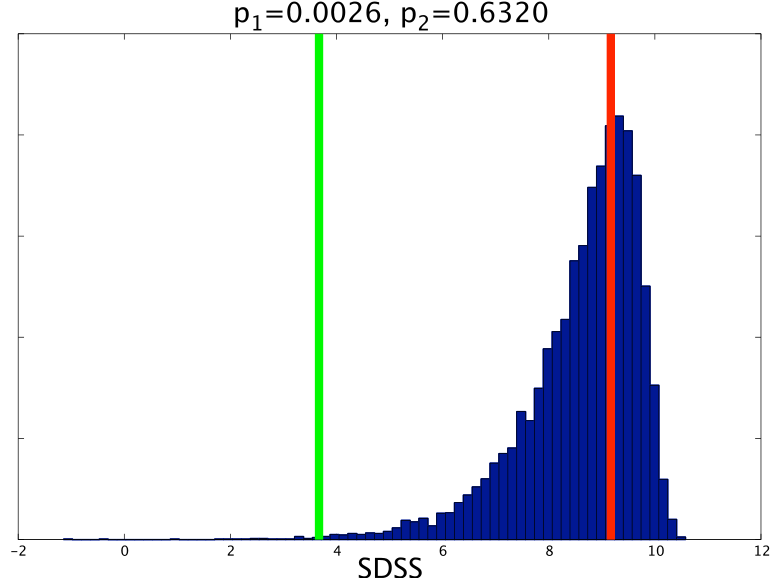}
\caption{As in Fig.~\ref{fig:snls-2sets}, but for the pairs
 CfA-1/SDSS and CfA-2/SDSS.\label{fig:sdss-2sets}}
\end{figure}

The results are shown in Figs~\ref{fig:snls-2sets} and
\ref{fig:sdss-2sets}.  It is clear that the CfA-1 subsurvey is
consistent with both the SNLS and SDSS surveys, whereas the CfA-2
subsurvey is inconsistent at very high significance in both cases.
This suggests that it is the selection of appropriate SNe, and not
necessarily the improved fitting method, that is key to achieving the
mutual consistency of the surveys within the JLA compilation

\section{Conclusions}\label{sec:conc}

SNe are now routinely used to constrain cosmological parameters, in
combination with CMB and large-scale structure observations. SNe
catalogues typically consist of SNe from a number of different
surveys, and most studies perform a joint analysis of all the surveys
within a given SNe catalogue. A key (often implicit) assumption
underlying these joint analyses is that the different SNe surveys that
make up the catalogue are mutually consistent. It is possible,
however, that unaccounted for systematics may cause tensions or even
inconsistencies in the parameters preferred by some surveys
comprising the catalogue. It is therefore extremely important to
establish whether any such inconsistencies exist before performing a
joint analysis.

In this paper we have performed a robust statistical test to look for
inconsistencies among the different surveys making up the UNION2 and
JLA catalogues, respectively. The test is based on a Bayesian
prescription, but has been adapted to be applied to analyses performed
using the standard $\chi^2$ analysis that is used most widely by the
SNe community. Our main finding is that, although the cosmological
constraints derived from the different surveys making up the UNION2
and JLA catalogues are consistent, the constraints on the parameters
associated with the SNe population, namely the multipliers $\alpha$
and $\beta$ of the stretch and colour corrections respectively, are
significantly different for the UNION2 catalogue, but consistent for
the JLA catalogue. The level of inconsistency exhibited by the UNION2
compilation suggests that one must exercise caution when interpreting
cosmological constraints derived from it with the usual joint
analysis.

\section*{Acknowledgements}\label{sec:ackn}

NVK thanks Rahman Amanullah for many interesting discussions regarding
supernova astronomy and data analysis. We thank Rick Kessler for
all his help with SNANA. FF is supported by a Research Fellowship from
Leverhulme and Newton Trusts. This work was performed primarily on
COSMOS VIII, an SGI Altix UV1000 supercomputer, funded by SGI/Intel,
HEFCE and PPARC. The work also utilized the Darwin Supercomputer of
the University of Cambridge High Performance Computing Service
(\texttt{http://www.hpc.cam.ac.uk}), provided by Dell Inc. using
Strategic Research Infrastructure Funding from the Higher Education
Funding Council for England.

\bibliographystyle{mn2e}
\bibliography{references}

\label{lastpage}
\end{document}